\documentclass[showpacs,preprintnumbers,amsmath,amssymb,nofootinbib]{revtex4}
\usepackage{graphicx}
\usepackage{epsfig}
\usepackage{bm}
\usepackage{amsfonts}
\newcommand{\Oo}{{\Delta_2}}

\newcommand{\eq}{\begin{eqnarray}}
\newcommand{\eqx}{\end{eqnarray}}
\newcommand{\ba}{\begin{equation}}
\newcommand{\ea}{\end{equation}}
\newcommand{\bal}{\begin{align}} 
\newcommand{\eal}{\end{align}}
\newcommand{\f}[2]{\frac{#1}{#2}}

\newcommand{\n}{\nonumber \\}

\newcommand{\dl}{\delta}

\newcommand{\Dl}{\Delta}

\newcommand{\bit}{\begin{itemize}} 
\newcommand{\eit}{\end{itemize}}

\def\la{\label}
\def\nn{\nonumber \\}
\def\bi{\bibitem}

\def\d{\partial}
\def\th{\theta}
\def\Del{\Delta}

\def\al{\alpha}

\def\be{\beta}

\def\va{\varphi}

\def\eps{\epsilon}

\def\la{\label}

\begin{document}

\title{Fourier-positivity constraints on QCD dipole models}

\author{Bertrand G. Giraud }
\author{ Robi Peschanski }
\affiliation{Institut de Physique Th\'eorique,\\
CEA, IPhT, F-91191 Gif-sur-Yvette, France\\
CNRS, URA 2306 }
\email{bertrand.giraud@cea.fr; robi.peschanski@cea.fr}

\today

\begin{abstract}
Fourier-positivity (${\cal F}$-positivity), i.e. the mathematical 
property that a function has a positive Fourier transform, 
can be used as a constraint on the parametrization of QCD dipole-target  
cross-sections or Wilson line correlators in transverse position space $r.$
They are Bessel transforms of positive transverse momentum dependent gluon 
distributions. Using mathematical ${\cal F}$-positivity constraints on 
the limit $r\to 0$ behavior of the dipole amplitudes, we identify the 
common origin of the violation of ${\cal F}$-positivity for various, however 
phenomenologically convenient, dipole models. It is due to the behavior 
$r^{2+\eps},\ \eps>0$ softer, even  slightly,  than color 
transparency.  ${\cal F}$-positivity seems thus to conflict with the 
present dipole formalism when it includes a QCD running coupling constant $\al(r)$. 
\end{abstract} 
\maketitle

\section{Origin of the problem}
\la{introd}

The introduction of the dipole formalism \cite{mueller} in 
QCD predictions for high-energy deep-inelastic scattering (DIS) and other 
processes at ``small $x$'' has prompted a 
lot of phenomenological activity with good success during the recent 
years. Initially, the dipole representation allowed one to express the DIS structure functions off a target 
 in terms of the (imaginary part of) the forward scattering amplitude ${\cal N}(Y,r)$ of a colorless
quark-antiquark  ($q\bar q$) pair on a target. 
 Here
$q\bar q$ 
is a colorless QCD dipole of transverse size $r,$ 
elastically scattering on the target at rapidity $Y.$ ${\cal N}(Y,r),$ up 
to a normalization, is the $q\bar q$-target total cross-section, and thus 
necessarily positive.

The  following Fourier-Bessel relations between transverse spatial and momentum coordinates have been written
 \cite{kovchegov},
\bal 
{\cal N}(Y,r) &= r^2 \int_0^\infty kdk\  
J_0(kr)\ 
\tilde {\cal N}(Y,k)\ ,
\nn 
 \tilde {\cal N}(Y,k)\ &=\  \int_0^\infty \frac {dr}r\  
J_0(kr)\ 
{\cal N}(Y,r)\ ,
\la{fpair}
\end{align}
where $\tilde {\cal N}(Y,k)$ is related to the $F_2$ 
structure function of DIS off the target. At that time, this connection  \cite{munpesch} implied that  $\tilde {\cal N}(Y,k)$ should be positive within a certain approximation. 
Within this approximation, \eqref{fpair} 
relate, through a Fourier-Bessel transformation, two physical observables 
which are both positive. This was our original 
motivation \cite{gipe} to discuss Fourier-positivity properties in general, hoping to get constraints on the QCD dipole models.
 
At present time and in a more general setting, it can be proven that QCD observables at small $x$ like ${{\cal N}(Y,r)}$ and $\tilde {\cal 
N}(Y,k)$ are expected to be  conjugated $positive$ functions through a 
Bessel transformation.  This characteristic feature of 
positive Fourier partners  is indeed more general in the dipole 
formalism and its various developments than the one that we originally 
considered \cite{gipe}. In \cite{gluon}, identical relations \eqref{fpair}  
are found for $q \bar q$ jet correlations in DIS. In that case, ${{\cal 
N}(Y,r)}$ is obtained from a specific correlator of Wilson lines and 
${\tilde {\cal N}(Y,k)}$ is the Weiszs\"acker-Williams 
transverse momentum dependent (TMD) gluon distribution of the target. 

Moreover, in Ref.\cite{marquet1} and in the previous
reference \cite{gluon}, a different TMD gluon distribution describes inclusive forward jet production and dijet 
correlations, respectively, in proton-nucleus ($p$-$A$) scattering. In that case, a similar but 
different Bessel relation can be written between this TMD gluon distribution and the total cross-section of the target with a gluon-gluon ($gg$) colorless dipole.  We will show, later on, that this relation between $p$-$A$ observables gives rise to the same constraints as \eqref{fpair}.

On a mathematical footing, we call  ``$\cal F$-positivity'' that mathematical property of a real positive function whose Fourier transform is itself 
positive. The aim of our paper is to derive some useful constraints on 
current QCD dipole (or related) models coming from  $\cal F$-positivity. 
More precisely, let us consider  the case of the relations \eqref
{fpair}.  Given a model parameterization for ${\cal N}(Y,r)\ge 0,$ we want 
to determine to which conditions it leads 
to a positive distribution $\tilde {\cal N}(Y,k)\ge 0.$ This problem 
arising for phenomenological QCD dipole models was already noticed in Refs.
\cite{gelis,marquet} and its relation to mathematical properties in \cite
{gipe,lappi}.

As a mathematical problem, $\cal F$-positivity is old and still not 
completely solved \cite{maths}. There exists  fundamental theorems \cite 
{bochner} ensuring the $\cal F$-positivity of real functions, as we 
shall see later on. However, up to 
our knowledge, there is no explicit property of the set of all 
$\cal F$-positive functions in 2-(or else) dimensions which would {\it a 
priori} suggest the parametrization of ${\cal N}(Y,r).$ In the 
absence of a definite practical scheme,  we recently developed \cite
{newfourier,diraccomb} a set of practical tools in the form of a 
hierarchy of necessary conditions. They hold for the general 
1-d and 2-d radial cases, which we will now try and apply specifically 
to the QCD dipole problem.

The plan of our paper is the following. In section \ref{matools}, we 
describe the mathematical derivation of the specific tools we will apply 
to the QCD dipole formalism. We insist in particular on the $\cal F$
-positivity constraints on the small-size limit of the dipole 
input. In section \ref{formul}, starting with the Fourier partners in Eq.\eqref{fpair},  we review examples from the main classes 
of phenomenologically useful dipole (or related) models and show how and why they often contradict $\cal F$-positivity. In section \ref{theory}, extending our study to the other type of Bessel relations \cite{marquet1,gluon},
 we show that, for both formulations,  the QCD dipole or Wilson line correlators models, 
including a running coupling constant,  violate   
$\cal F$-positivity for the same reason, which was yet unknown. The final section \ref{summary} summarizes our 
results and discusses the origin of the $\cal F$-positivity failure, opening 
possible ways out to reconcile an improved dipole formalism with  ${\cal F}$
-positivity.

\section{Mathematical aspects and tools}
\la{matools}
The fundamental mathematical property characterizing $\cal F$-positivity
uses the Bochner theorem \cite{bochner}. Applied for instance to the QCD dipole 
problem issued from relations \eqref{fpair}, it can be expressed as follows. $\cal F$-positivity of 
$\tilde{\cal N}(Y,k)$  is equivalent to the statement that  ${{\cal 
N}(Y,r)}/{r^2}$ is not
only positive but also {\it positive-definite}. In the full 
2-dimensional transverse coordinate space, positive-definiteness means 
that for any set of positions in ${\mathbb R}^2$, $\{{\vec r_i}, 
i=1,...,n\}$, for any $n \in {\mathbb N}$ and for any set of numbers 
$\{{ u_i}, i=1,...,n\},$ the $n \times n$ matrix $\mathbb M$ with 
elements 
${{\cal N}(Y,|\vec{r_i}\!-\!\vec{r_j}|)}/{|\vec{r_i}\!-\!\vec{r_j}|^2})$ 
is positive definite. Namely, 
\ba
\sum_{i,j=1}^{n}u_i\ \ \frac{{\cal N}(Y,|\vec{r_i}\!-\!\vec{r_j}|)}{|\vec{r_i}\!-\!\vec{r_j}|^2}\ \ u_j\ >\ 0\ ,\quad 
\forall  u_i,\ \forall  \vec r_i,\
\forall n\ . 
\la{positive}
\ea
In other terms, the lowest eigenvalue of $\mathbb M$
 remains positive for all $\vec r_i, u_i$ and all values of $n.$

The Bochner theorem with its applications appears to be still the major
tool in 
the domain. However, to our knowledge, there does not yet exist a
mathematical
 classification of Fourier-positive  functions which, for instance,
could constrain at once
 an appropriate parametrization for model building. Testing 
positive-definiteness  \eqref{positive} cannot be done concretely, due
to the infinite 
number of the constraints. Conversely, a calculation of 
Fourier transforms for checking Fourier-positivity is obviously 
possible, but this does not give  easy means to 
select 
{\it a priori} appropriate $\cal F$-positive sets of functions, or to find the origin of their eventual $\cal F$-positivity violation.

Our approach is thus to find new constraints of Fourier-positivity 
allowing for 
simple and efficient selection rules on the QCD model parametrizations 
starting with both $positive$ and $positive-definite$ input functions
whose  ${\cal N}(Y,r)/r^2$ in Eq.\eqref{1} is an example.

In our previous papers \cite{gipe,newfourier}, we performed a study of 
 large sets of $positive$ and $positive-definite$ functions 
in 1 and 2 (radial) dimensions. We used bases of positive 
polynomials or convexity properties. We found some intricate 
features of the problem. 
 
In the present work we will use tools based on the Bochner theorem, coming from Ref.\cite{diraccomb} where  a satisfactory 
detection 
of Fourier positivity, thoroughly tested for large sets of functions in 
one- 
and radial two dimensions, was obtained. Let us sketch our formulation in the specific case of the (radial) 2-dimensional problem, appropriate for the present study.

{\it Matrices $\mathbb M$ from a point lattice.}
The  tool we will use is the set of necessary conditions coming from the
positive-definiteness condition \eqref{positive} for a discrete 
2d-lattice of points, $\vec r_i=\{n_i^{(1)}\!  r,n_i^{(2)}\!  r\},\ n_i^{(1),(2)} \in {1,\cdots,n}.$
 For example, and futher application, let us consider the $3\times3$ matrix 
 \ba
\{{\mathbb M}_{i,j}\}{_{_3}}\ \equiv\  \left\{\, 
\frac{{\cal N}(Y,|\vec{r_i}\!-\!\vec{r_j}|)}{|\vec{r_i}\!-\!\vec{r_j}|^2}
\,\right\}
,\quad \vec {r_i} \ =\ \{0,0\}\ ,\ \{0,r\}\ ,\ \{r,0\}\ ,
\la{3-lattice}
\ea
 which, denoting for simplicity $\psi(r)\equiv {{\cal N}(Y,r)}/{r^2},$ leads to the $\cal F$-positivity conditions for the matrix
\ba
\begin{pmatrix}
\psi(0)& \psi(r)& \psi(r\sqrt 2)\\
\psi(r)& \psi(0)& \psi(r)\\
 \psi(r\sqrt 2)& \psi(r)&\psi(0)\\ 
\end{pmatrix}\ .
\la{3x3matrix}
\ea
Positive-definiteness implies positivity of the matrix determinant
and of its minors along its diagonal, leading, up to a rescaling of $r,$ to the inequalities
\ba
\psi(0)\ > \  
\psi(r)\  > 
\  2\ \frac {\psi^2(r/\sqrt{2})}{\psi(0)}
-\psi(0)\ .
\la{fptest}
\ea
Note that larger lattices including the points of \eqref{3-lattice} 
will always lead to
condition \eqref{fptest} and will add others \cite{diraccomb}, forming a hierarchy of necessary conditions for $\cal F$-positivity.

An important addendum to the $\cal F$-positivity tests for a 2-dimensional radial function $\psi(r),$ as noticed in our Ref.\cite{newfourier}, is that they can be extended to the action of the radial Laplacian
\ba
\Oo \left[\psi\right](r)\ \equiv\ - \f 1r\f d{dr}\left(r\f d{dr}\right)
\left[\psi\right](r)\quad \Rightarrow\quad\Oo\left[{\cal N}/{r^2}\right](Y,r) =  \int_0^\infty k^3dk\  
J_0(kr)\ \tilde {\cal N}(Y,k)\, .
\la{operatorJ}
\ea
Hence ${\cal F}$-positivity applies also for $\Oo \left[\psi\right],$ as well as for its iterations, provided the integral in \eqref{operatorJ} converges.

\section{${\cal F}$-positivity tests of the QCD dipole amplitude}
\la{formul}

We shall now apply ${\cal F}$-positivity tests to the  
 amplitude parametrizations of ${\cal N}(Y,r).$  For this we 
review various popular classes of  models used 
in the phenomenology of low $x$ processes.

\subsection{BFKL-type models}

A phenomenologically successful model 
was proposed by \cite{munier}. It is based upon an effective BFKL-type 
amplitude  corrected at large transverse size for unitarity. It 
reads
\begin{align}
 {\cal N}(Y,r) &= { {\cal N}_0}\ 
\exp{
\left\{2\left(\gamma_s+\frac{\log(2/rQ)}{\kappa \lambda Y}\right)\ \log\left(\frac {rQ}2\right)\right\}
}
\quad {\rm for}\quad rQ \le  2,\nn 
{{\cal N}}(Y,r) &= 1 - \exp{\left\{-a \log^2(b  rQ)\right\}}\quad {\rm for}\quad rQ > 2\ .
\la{IIM}
\end{align}
$Q \equiv Q(Y)$ is the rapidity-dependent ``saturation scale'' and 
$\gamma_s,\kappa, \lambda$ are characteristic constants of the BFKL 
amplitude at leading log level \cite{munier}. Figure \ref{1} shows the  
functions $\cal N$
and $\tilde N$ in the appropriate scaled units, $rQ$ and $k/Q$ 
respectively, for physical values of $Y =4,6,8.$ The left-hand plot 
displays the dipole amplitude together with its value divided by 
$(rQ)^2.$ The right-hand plot is the gluon amplitude in log-log plot, 
exhibiting zeroes, showing explicitely the ${\cal F}$-positivity 
violation of the BFKL-inspired dipole amplitude.
\begin{figure}[htbp] 
\includegraphics[width=6cm]{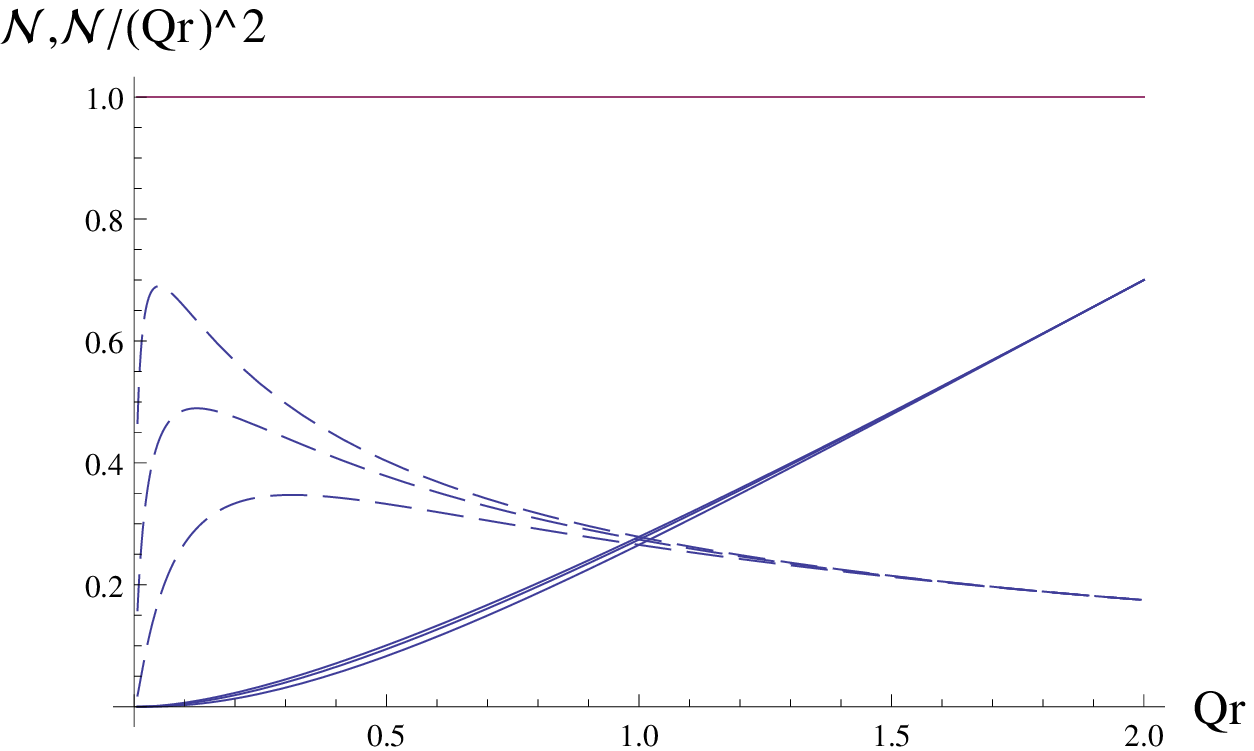} 
\hspace{2cm}
\includegraphics[width=6cm]{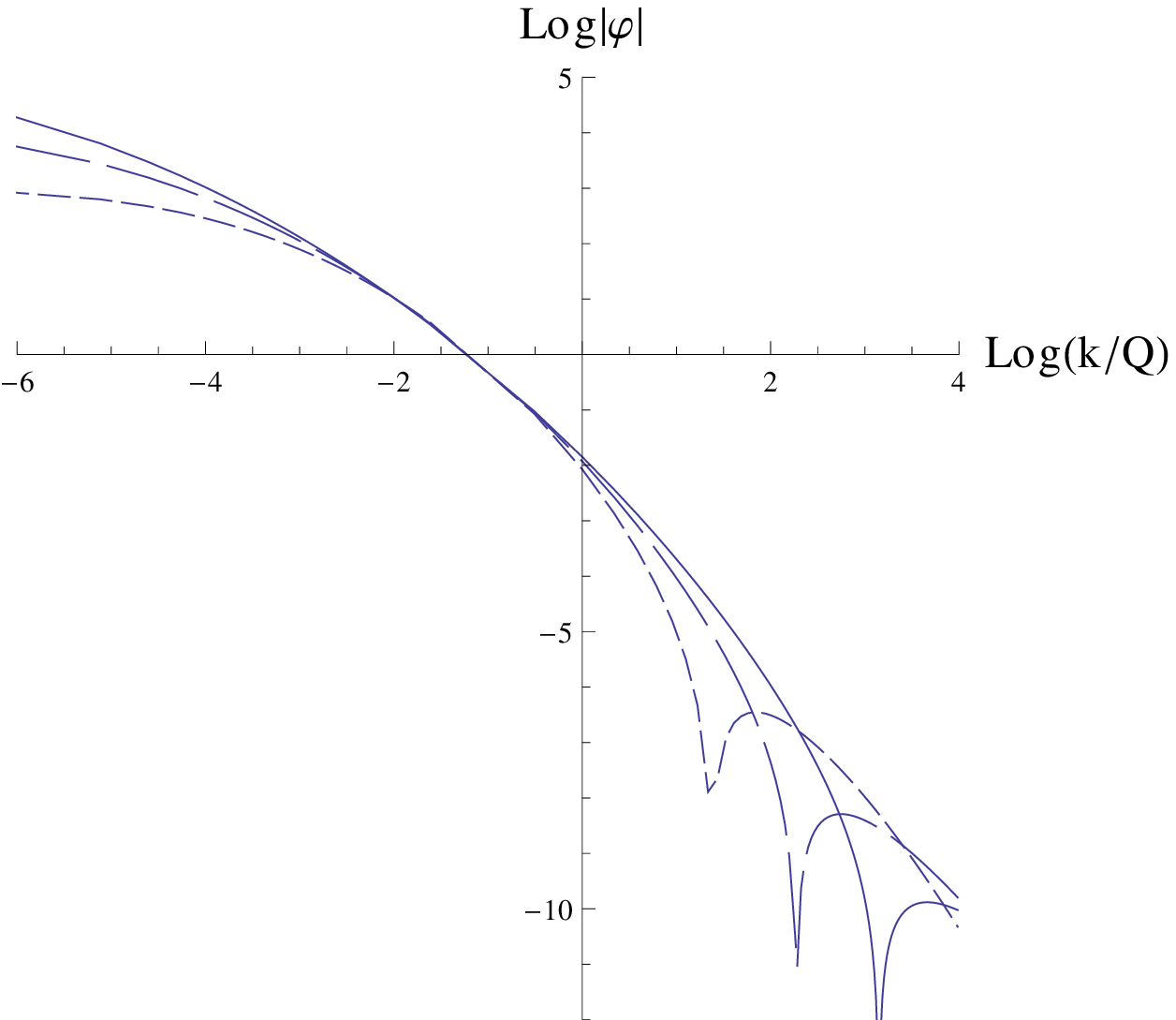}
\caption{{\it ${\cal F}$-positivity tests of the BFKL-inspired dipole 
amplitude.} Left, the dipole amplitude:  $full\ line,$ ${\cal N}(Y,r),$ $dashed\ line,$ $ {\cal 
N}(Y,r)/(rQ)^2,$ as a function of $rQ.$ {Right,} its Fourier-transform $\va(Y,k/Q) \equiv \tilde 
{\cal N}(Y,k)$ in a $\log |\va|,\log (k/Q)$ plot for  $Y=4,6,8$. The dips signal the zeroes of $\va$ after which it is negative. The range of $Y$ 
runs from bottom to top in the left plot, and from right to left in the 
right plot. }
\la{1}
\end{figure}
A characteristic feature that we retain from  the function  $ {\cal 
N}(Y,r)/(rQ)^2,$ which forms the Fourier pair with the 
distribution $\va(Y,k/Q) \equiv \tilde {\cal N}(Y,k)$ is that it increases 
from zero at small $r.$ This is a signature of ${\cal F}$-positivity 
violation since the constraint  \eqref{fptest} is not satisfied. Then, 
already at the first level of our hierarchy of constraints, the amplitude
\eqref{IIM} does not obey ${\cal F}$-positivity. This can be traced 
back to  the logarithm $\log(2/rQ)$ in the exponent of the BFKL-inspired expression \eqref
{IIM},  since it drives the amplitude to zero when $r \to 0,$ for any 
value of the parameters.  In particular, the violation is not related to the presence of non-regular behavior at the transition point $Qr=2,$ as sometimes conjectured.

One may object that when this logarithm dominates, the effective  BFKL amplitude is no more justified. However, as we now will see, ${\cal F}$-positivity violation due to the small-$r$ behavior of the amplitude is not restricted to BFKL-inspired models.

\subsection{Golec-Biernat-Wusth\"off-type models}
\la{gbw}
The Golec-Biernat-Wusth\"off model \cite{golec} (GBW) is a quite popular 
model of the dipole amplitude. In a recent paper \cite{iancu} a generalized expression of this type has 
been taken for the initial input amplitude for a 
phenomenological application of the Balitsky-Kovchegov (BK) evolution equation \cite{balitsky} with rapidity. It reads
\begin{align}
 {\cal N}(Y,r) = \left\{
1-\exp{
\left[
-\left(
{\frac 14}(rQ(Y))^{2}
\right)^{ p}
\right]}
\right\}^{{1/{ p}}}\ ,
\la{GBW}
\end{align}
where $p=1$ for the original model \cite{golec} while $p>1$ is clearly 
preferred in the recent application \cite{iancu}. One sees 
from Fig.\ref{2} that  ${\cal F}$-positivity is still violated even if  
the inequalities \eqref{fptest} are satisfied. The function $ {\cal 
N}/(rQ)^2,$ appears to be 
slightly decreasing with $Qr.$
\begin{figure}[htbp] 
\includegraphics[width=6cm]{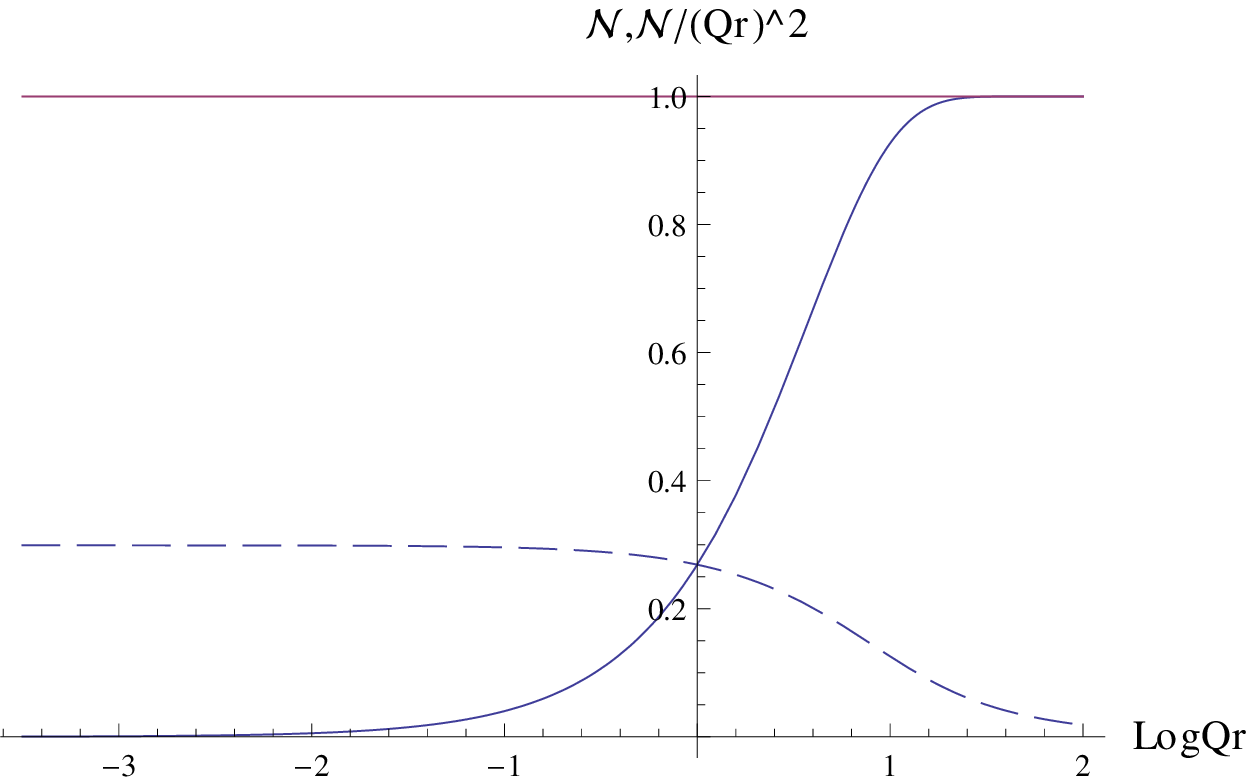} 
\hspace{2cm}
\includegraphics[width=6cm]{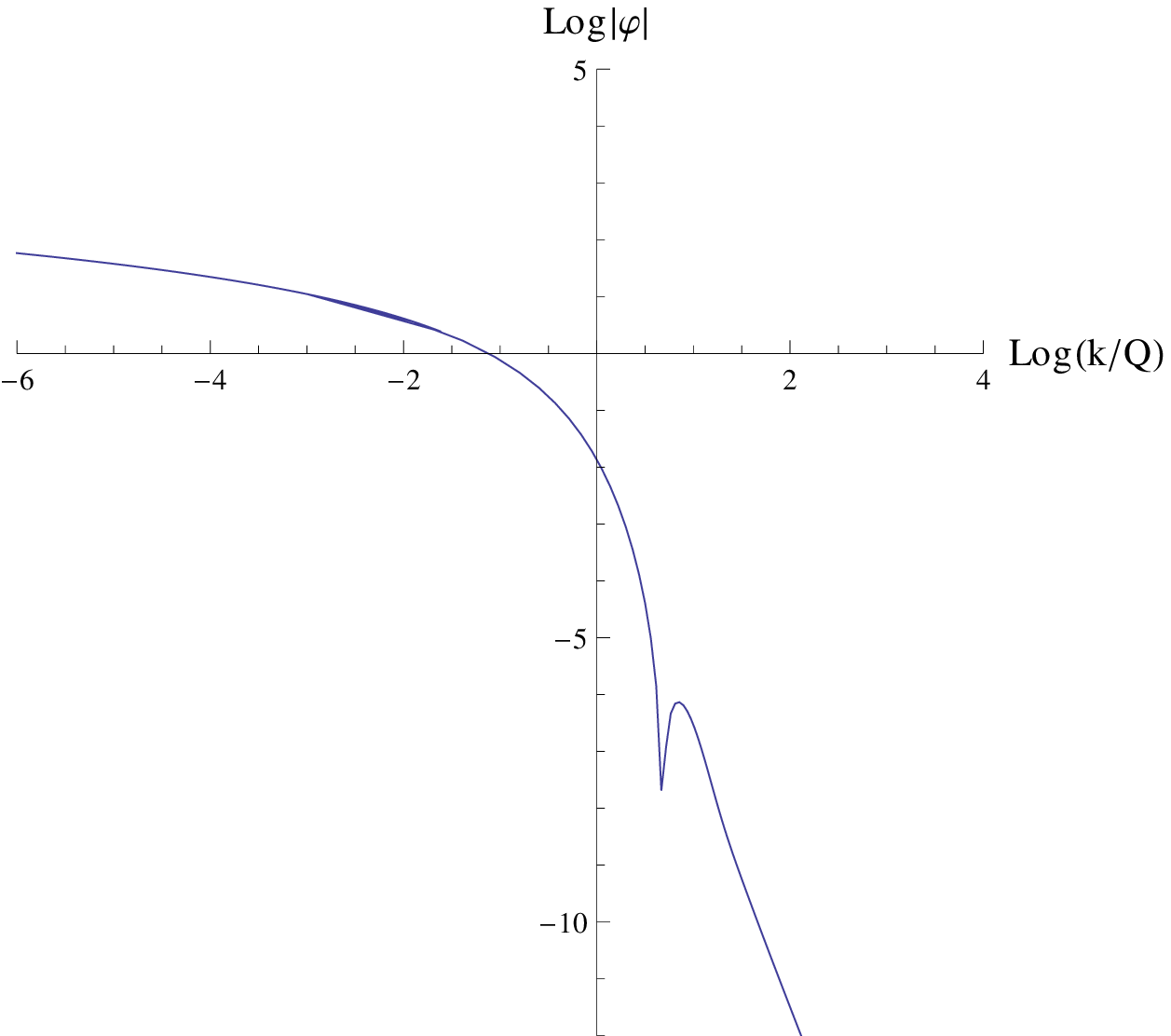}
\caption{{\it ${\cal F}$-positivity tests of the GBW-type dipole 
amplitude.} 
Left, the dipole amplitude:   ${\cal N}(Y,r,)$ $full\ line,$  $ {\cal 
N}(Y,r)/(rQ)^2,$ $dashed\ line,$ in a semi-log plot as a function of $\log (Qr).$  {Right,} its Fourier-transformed $\va(Y,k/Q) \equiv \tilde 
{\cal N}(Y,r)$   in a $\log|\va|,\log(k/Q)$ plot. The dip corresponds to the first zero of $\va.$ The Fourier pair of 
functions  is displayed for {p=1.148} in Eq.\eqref{GBW}, as found in Ref. \cite{iancu}, table 1.}
\la{2}
\end{figure}
The situation clarifies if one considers the radial Laplacian operator $\Oo$ defined in \eqref{operatorJ}. One finds 
at small $r$
\ba
\Oo [\psi](r)\ = \ - \left\{\f {d^2}{(dr)^2}+\f 
1r\f d{dr}\right\}\left[\f{{\cal N}}{r^2}\right](Y,r)\ \approx \ - \left\{\f {d^2}{(dr)^2}+\f 
1r\f d{dr}\right\} \left[\f 14-\f 1{8p}\left(\f {r^2}{4}
\right)^p\right]\ =\ 
\f {2p }{4^{p+1}}\  r^{2(p-1)} \, .
\la{DtestGBW}
\ea
Alongside Eq.\eqref{fptest}, ${\cal F}$-positivity requires also $\Oo 
[\psi](0)>\Oo [\psi](r),$ which is satisfied only for $p<1$ in Eq.\eqref
{DtestGBW}. Hence, any function with $p>1$   appearing for phenomenological preference in \cite{iancu} does not satisfy 
the ${\cal F}$-positivity constraint.

\subsection{Mc Lerran-Venugopalan-type models}
\la{rmv}

The Mc Lerran-Venugopalan model \cite{MV} (MV) for the dipole amplitude 
is well known in low $x$ phenomenology and theory. It gives rise to 
various recent versions of MV-type, with successful applications to 
phenomenology. For our present study we will retain first variations of 
the original model used as a phenomenological input for $Y$-dependence 
\cite{lappi1}, namely
\ba
 {\cal N}(Y,r) = 1-\exp{\left[-{{\frac 14}}(rQ(Y))^{2\gamma}\log\left(e\  e_c+\f 1{r\Lambda}\right)\right]}\ ,
\la{mv}
\ea
where the constants $(\gamma, e_c$) allow a best fit search while they 
are equal to $(1,1)$ in the original MV model. Using the condition 
\eqref{fptest}  leads to impose $\gamma \le 1.$ This 
is satisfied by the  MV formulas \eqref{mv} with $\gamma=1$ (including the model with $e_c \ne 1$ \cite{lappi1}) but not satisfied for the so-called AAMQS 
model \cite{AAMQS} which has $\gamma = 1.119.$

It is worthwhile to note that a problem of $\cal F$-positivity violation related to the 
MV model has been noticed in Ref.\cite{marquet} when one uses the 
original version instead of the approximation by the logarithm in \eqref
{mv}. It  contains  a cut-off  $\theta(r\Lambda).$ $\cal F$-positivity 
violation in this case is thus well detected by our test \eqref 
{fptest} since $0=\psi(0)< \psi(r)$ for r larger than the 
cut-off. $\cal F$-positivity is restored \cite{marquet} by replacing the cut-off by
a smooth extrapolation. Indeed, the corresponding derivative of $\va(r)$ at $r=0$ is negative, 
ensuring $\psi(0)> \psi(r)$ as required by $\cal F$-positivity. In this case, this criterion is indeed sufficient.

One also considered recently \cite{iancu} the MV model with running coupling constant, namely \cite{MV},
\begin{align}
 {\cal N}(Y,r) = \left\{
1-\exp{
\left[
-\left(
{\frac 14}(rQ(Y))^{2}
{\bar \alpha(rC)}
\left[
1+\log\left(\frac{\bar \alpha_{sat}}{{\bar \alpha(rC)}}\right)
\right]
\right)^{ p}
\right]}
\right\}^{{1/p}}\ ,
\end{align}
where ${\bar \alpha(rC)}\propto {\log^{-1}(1/(rC)}$ is the running QCD 
coupling in transverse coordinate space and $C,\alpha_{sat},p$ are 
phenomenological constants. 
It is easy to realize that
$ {\cal N}(Y,r)/r^{2}\approx 
\left[
\frac{\log\log \frac 1r}{\log \frac 1r}
\right] \to 0$ when $ r\!\to\! 0:$
$\cal F$-positivity is thus violated, in a kind of ``weak'' logarithmic 
form, due to the running coupling. We shall see now that this feature is 
not subsidiary; actually it is related to a deeper theoretical problem 
of the QCD dipole formalism with a running coupling constant $\al(r).$

\subsection{Saturation model with DGLAP evolution}
\la{gbbk}

On a 
theoretical ground, when $r\to 0,$ a correspondence with large 
$k/Q$ of the observable, the QCD dipole model is faced with the problem 
of compatibility  with the DGLAP evolution equations. Let us examine 
which constraints this generates on  
$\cal F$-positivity tests. For this sake, and in order to take more 
general lessons, we shall consider first a well-known model \cite{msat} 
introducing a modification of the saturation model in order to take into 
account DGLAP evolution. It reads
\ba
{\cal N}(Y,r) = 
1-\exp{
\left(
-\frac 1{3\sigma_0}\pi^2r^{2}
\bar \alpha(\mu^2) \times xg(x,\mu^2)\right)
}\ ,\quad x\equiv e^{-Y}\ .
\la{satdglap}
\ea
Here  $xg(x,\mu^2)$ is the gluon distribution function in the proton considered at 
momentum fraction $x$ and  $r$-dependent  scale $\mu^2 =  C/{r^2}+\mu_0^2,$
 with $\alpha(\mu^2)\propto \log \Lambda_{QCD}^2/\mu^2$ 
and $\sigma_0, C$  phenomenological 
constants fitted to the deep-inelastic data.

At small $x,$ the leading behavior of the gluon distribution function can be obtained from the resummation of the double leading logarihtms, namely
\ba
xg(x,\mu^2) \approx 
\sum_n 
\frac{\left[\ \log(1/x) \int_{\mu_0^2}^{\mu^2}\!\! \alpha(k^2)\  {dk^2}/{k^2}\ \right]^n}{(n!)^2} \sim 
\exp{\sqrt{\log\f 1x\ \log\log\f{{\mu}^2}{{\mu_0}^2}}}\ ,
\la{g}
\ea
leading to
\begin{align}
 {\cal N}(Y,r)\ \approx\ r^{2}\ \frac 1{\log  \frac1{r^2}}\ 
\exp{\sqrt{Y\ \log\log{\frac 1{r^2}}}}\ \to\ 0\quad {\rm when}\ r\!\to\! 0\ .
\la{limit}
\end{align}
In this example, we see that the double leading log resummation of the 
perturbative expansion of the gluon distribution function is unable to fully 
compensate for the running of the coupling constant in front of \eqref
{limit}.

There is a model-independent lesson to be remembered from the example \eqref{g} and the 
investigation of the previous subsection: The QCD asymptotic freedom, appearing in the dipole 
amplitude in the form of an overall running coupling  going logarithmically to zero with the 
dipole size seems to systematically contradict $\cal F$-positivity at large enough 
transverse momentum. The inverse logarithmic behavior of the running coupling constant is only partially compensated by the resummation at large rapidity $Y$ of the perturbative expansion of the dipole amplitude.

\section{$\cal F$-positivity: a generic problem.}
\la{theory}

The various phenomenological models discussed in the previous sections show that 
there often exists a common problem with $\cal F$-positivity at large values of $k/Q,$ 
i.e. when the gluon momentum $k$ is large w.r.t. the typical (saturation) 
scale $Q$ of the target. We have seen that the problem is due 
in all cases to the limit behavior of the amplitude ${\cal N}(Y,r)$ when  
 $r\to 0.$ As shown in subsections \ref{rmv} and \ref{gbbk},  when the running $\al(r)$
of the QCD coupling constant  is taken into account, it is responsible 
for the violation of the $\cal F$-positivity test \eqref{fptest}. 
 Let us extend now the results obtained with the Fourier partners of 
\eqref{fpair} to the other type of Bessel relation appearing in  Refs \cite
{marquet1,gluon}. It relates the $gg$ dipole-target forward amplitude 
$T(Y,r)$ to the TMD gluon distribution $\tilde {\cal T}(Y,k)$. The Bessel 
relation in those cases reads
\bal 
 \tilde {\cal T}(Y,k)\ =\  \int r{dr}\  
J_0(kr)\ 
\f 1r\frac {\d}{\d r}\ \left(r\frac {\d}{\d r}{\cal T}(Y,r)\right)\ >\ 0\ ,
\la{fpair0}
\end{align}
giving rise again to a $\cal F$-positivity constraint.

To analyze this constraint, we introduce a generic parameterization  of the $gg$ dipole-target forward when $r\to 0,$ namely, 
\ba
{\cal T}(Y,r)\ \propto\ r^{2+\eps} \quad {\rm when}  \quad r\to 0\ .
\la{transparency}
\ea
``Color transparency'' corresponds to $\eps=0.$ Choosing  $0 < \eps \ll 1$
 allows one to provide an effective description of the effect of the running QCD coupling 
constant only partially compensated by resummation effects. As we shall now derive, this choice leads to a violation of $\cal F$-positivity.

Note that the same  behavior \eqref{fpair0} applied to
 $N(Y,r)$ in \eqref{fpair} leads also to the  violation of  $\cal F$
-positivity. Indeed one finds  
$N(Y,r)/r^2 \sim r^{\eps},$ increasing with $r$ and thus violating the constraint \eqref{fptest}. 

 In the integrand of the Bessel transform \eqref{fpair0}, one recognizes the opposite of the Laplacian operator of  \eqref{operatorJ}, namely,
\ba
 \f 1r\f d{dr}\left(r\f d{dr}{\cal T}(Y,r)\right)\ =\ -\Oo \left[{\cal T}\right](Y,r)\ \sim \ (2+\eps)^2\ r^{\eps}\quad {\rm when}  \quad r\to 0\ .
\la{proof0}
\ea
Hence, for $\eps >0,$ the constraint \eqref{fptest} and thus $\cal F$-positivity are violated.

To summarize our discussion, the running $\al(r)$ of the QCD coupling constant in front of the dipole amplitude ${\cal N}(Y,r)$ or the Wilson line correlator  ${\cal T}(Y,r)$   lead to a violation of $\cal F$-positivity tests in relation with the small-$r$ behavior. It corresponds to a small transverse distance between the Wilson lines or between the two gluons in the $gg$-dipole, respectively.

\section{Summary and outlook}
\la{summary}

We have discussed the constraints on QCD dipole models (and its extensions) arising from $\cal F$-positivity. Inded, the dipole cross-section or the Wilson line correlators in transverse position space are related through Bessel transformations \cite{gluon} to  two different transverse momentum dependent gluon distributions. These can be directly expressed in terms of observables and thus required  to be positive. The full 
mathematical  characterization of  $\cal F$-positive functions is  still an 
open problem. However, in Refs. \cite{gipe,newfourier,diraccomb}, we have proposed a set of 
rather simple conditions for  $\cal F$-positivity. In the present work, we show that they provide quite stringent necessary constraints on QCD dipole models, if not on the dipole formalism itself. Let us quote the main results:

i) Using a set of $\cal F$-positivity constraints \cite{newfourier,diraccomb} based on the  Bochner theorem \cite{bochner}, we show that dipole models, frequently used in low $x$ physics phenomenology, often do not satisfy $\cal F$-positivity.  We show that this violation is related to the limit behavior of the model amplitudes when the transverse size $r\to 0.$

ii) In all cases\footnote{In subsection \ref{gbw} a higher rank convexity constraint \eqref{DtestGBW} is violated.} when $\cal F$-positivity is violated, this happens if the behavior of the dipole amplitudes or Wilson line correlators is $\sim r^{2+\eps},\ 0<\eps\ll 1$ when the transverse size $r \to 0,$
where ``color transparency'' corresponds to $\eps=0.$ 

iii) On a more theoretical level, $\cal F$-positivity appears to put a question mark on the QCD dipole (or Wilson correlators) formalism itself. Indeed, when considering a dipole model with a coupling constant running with $r,$ it seems difficult to satisfy the constraint \eqref{fptest}. Indeed, asymptotic freedom  of $\al(r)\sim 1/\log(1/r)$ imposes a decreasing behavior of the amplitude faster than $r^2$ when $r \to 0.$ It appears only partially compensated by the all-order perturbative  resummation of the double logarithms in the low $x$ domain.

To our knowledge, there does not yet exist a satisfactory solution of the  $\cal F$-positivity constraints on the QCD dipole formalism. One way out of the difficulties would be that the Bessel transform relations could be relaxed at next-leading orders. However then, the question arises how to connect the $r$-dependent amplitudes with the transverse momentum observables. Another solution \cite{iancu2}, would be to modify the Wilson lines formalism starting with the $k$-dependent running coupling in order to preserve $\cal F$-positivity. Indeed, it is conceivable that the inner structure of a QCD dipole, which is after all a composite system of two quarks or two gluons, is resolved at very high transverse momentum of the probe.

Hence the  $\cal F$-positivity problem of the QCD dipole formalism seems to  deserve more study, both from phenomenological and theoretical points of view. 

\section{Acknowledgements}
We want to thank Edmond Iancu, Cyrille Marquet and Gregory Soyez for stimulating discussions.

\end{document}